\documentstyle[twoside,epsf,12pt,a4]{article}
\catcode`\@=11 \relax
\def\ps@burst{%
        \def\@oddhead{}  \def\@evenhead{}
        \def\@oddfoot{%
{\sl GSI-Preprint 95-45\hfil}
        } 
        \let\@evenfoot\@oddfoot
} 
\catcode`\@=12 \relax

\begin{document}
\thispagestyle{burst}
\begin{center}
{\Large\bf
Instabilities in nuclei
}
\end{center}

\vspace*{9mm}

\hspace*{9mm}\begin{minipage}[t]{12cm}
{\large
L.P. Csernai$^{1^\dagger ,2}$, J. N\'emeth$^3$ and
G. Papp$^{3^\dagger ,4}$
}\\[5mm]
$^1$Section for Theoretical Physics, Department of Physics\\
\hbox{\,\,\,}University of Bergen, All\'egaten 55, N-5007 Bergen, Norway,\\
$^2$Theoretical Physics Institute, University of Minnesota\\
\hbox{\,\,\,}Minneapolis, Minnesota 55455, USA\\
$^3$Department of Theoretical Physics, L. E\"otv\"os University,\\
\hbox{\,\,\,}H-1088 Budapest, Hungary\\
$^4$Gesellschaft f\"ur Schwerionenforschung, \\
\hbox{\,\,\,}D-64220, Darmstadt, Germany\\

\vspace*{6mm}{\it

Submitted to {\it Heavy Ion Physics}
}\\[9mm]
    \hrule height 0.4pt
    \vskip 8pt

{\bf Abstract.}
The evolution of dynamical perturbations is examined in nuclear
multifragmentation in the frame of Vlasov equation. Both plane wave and bubble
type of perturbations are investigated in the presence of surface (Yukawa)
forces. An energy condition is given for the allowed type of instabilities and
the time scale of the exponential growth of the instabilities is calculated.
The results are compared to the mechanical spinodal region predictions.
\\[9mm]
    \vskip 10pt
    \hrule height 0.4pt
\end{minipage}

\newpage

\section*{\large\bf 1. Introduction}
\label{sec-intr}

Let us consider a spherically expanding nuclear system in  the
metastable
nuclear fluid phase when it reaches the freeze-out at time $\tau_{fr}$.
Although at the freeze-out the fermionic degrees of freedom are frozen-out,
and internucleon collisions cease, softer long range nuclear interactions are
still effective, and represented by a nuclear mean field potential,
$U(\vec{r})$.

We will assume that the system, both before and after freeze-out
undergoes a spherical, scaling expansion.  Such an expansion can be
represented by a four-velocity field, $u_\mu = x^\mu / \tau$, where
$\tau = \sqrt{ t^2 - x^2 - y^2 -z^2 }$,  for the internal regions
of our expanding system (but not for the external surface).
This flow pattern is invariant under
Lorentz transformation, i.e., the points of the interior of our expanding
system are physically identical and indistinguishable from one another.
Consequently in the interior, in the Local Rest (LR) frame all thermodynamical
and fluid-dynamical quantities are equal, and from the point of view of
instabilities all internal points are equivalent.  We will exploit these
symmetry features although in the calculations we will use a non-relativistic
approximation.

  Let us assume that the nucleon phase space distribution
before, and at the freeze-out,  $\tau_{fr}$, is a Fermi distribution:
\begin{equation}
f_0(\tau_{fr},\vec{r},\vec{p}) =
C \left\{ 1 +
       \exp\left[\frac{[\vec{p} - m \vec{r}/\tau_{fr}]^2}{2mT_{fr}} -
            \frac{\mu_{fr}}{T_{fr}}\right]
\right\}^{-1}
\label{e:fd}
\end{equation}
where this form assumes a correlation between the momentum and radial
distribution arising from radial expansion, $C=g/(2\pi \hbar)^3$ is the
normalization, and $g$ is the degeneracy of nucleons, so that
the proper (LR) density is
$n_0(\tau_{fr},\vec{r})=\int d^3p\ f_0(\tau_{fr},\vec{r},\vec{p})$.
We assume that in the interior of the collision zone the  freeze-out
density is constant: $n_0(\tau_{fr},\vec{r}) = n_{fr}$.
In the center of the collision zone this is an
ideal Fermi distribution, while at finite radii,
$|\vec{r}\hspace*{2pt}|$, the distribution is
boosted (using non-relativistic, Galilei transformation), with a radially
directed and radially linearly increasing flow velocity of $\vec{v} =
\vec{r}/\tau$.   We can also introduce the LR momentum:
$\vec{P}(\tau , \vec{r}\hspace*{2pt}) = \vec{p} - m \vec{r}/\tau$.

Furthermore, let us assume that after the freeze-out for $\tau>\tau_{fr}$,
the system expands homogeneously according to the
collisionless Vlasov equation.  The distribution function, $f$, is the
solution of the Vlasov equation with a mean-field
potential, $U(\vec{r})$,
\begin{equation}
\frac{\partial f}{\partial \tau} + \frac{ \vec{p} }{ m }
\frac{\partial f}{\partial \vec{r} } -
\frac{\partial U}{\partial \vec{r}}\
             \frac{\partial f}{\partial \vec{p}}\  = 0 .
\label{e:vl}
\end{equation}
In the special case of a homogeneous system, where the last term vanishes, for
such a free coasting expansion in the local rest frame is just obtained by
replacing $\vec{p}$ by $\tau \vec{p} / \tau_{fr}$ in Eq. (\ref{e:fd})
\cite{csm95,csmm95}:
\begin{equation}
f_0(\tau,\vec{r},\vec{p}) =
C \left\{ 1 +
       \exp\left[\frac{[\vec{p} - m \vec{r}/\tau]^2}{2\ m\ T_{eff}} -
            \frac{\mu_{eff}}{T_{eff}}\right]
\right\}^{-1}
\label{e:f0}
\end{equation}
where
$T_{eff} = T_{fr} (\tau_{fr}/\tau)^2$ and $\mu_{eff} =
\mu_{fr} (\tau_{fr}/\tau)^2$.
The condition, that the ratio of the chemical potential and the temperature
is constant during the expansion, in a usual thermodynamical system,
corresponds to an adiabatic process.
The density of the system changes with time in this inertial expansion
as $n_0(\tau) = n_{fr} (\tau_{fr}/\tau)^3$.  This solution is valid starting
from the freeze-out, $\tau_{fr}$, and until inhomogeneities will spontaneously
develop in the system at some threshold time, $\tau_{th}$. Before this time
small perturbations will smooth out due to the mean field potential, while
after this threshold time the density dependent mean field will enhance
fluctuations.  So after this threshold time the density will not be
homogeneous any more.

Note that this post freeze-out distribution, $f$,
is not a thermal equilibrium distribution
function, and the effective parameters, $T_{eff}$ and $\mu_{eff}$ are just
carrying the memory of the last equilibrium thermal parameters, $T_{fr}$ and
$\mu_{fr}$, but these are not the usual thermodynamical parameters.  This can
be
easily seen if the expansion is not spherically symmetric
\cite{csm95}.

If we would have
a thermal expansion following $\tau_{fr}$, the Equation of State (EOS) would
determine the time dependence of the physical temperature in an adiabatic
expansion. Generally this would not coincide with $T_{eff} = T_{fr}
(\tau_{fr} /\tau)^2$,
only in the case of a large system, where the flow dominates the energy.
For example if the EOS is that of an ideal Stefan-Boltzmann
gas, ($\partial p /\partial e = c_0^2$, $c_0^2 = 1/3$ and $e = c T^4$), then
$T(\tau) = T_{fr} (\tau/\tau_{fr})^{-3c_0^2}$, which differs from $T_{eff}$.

We study the
stability of the system and the occurrence of instabilities arising from
the mean field. Such an instability may lead to a rapid multifragmentation
of our system.

\section*{\large\bf 2. Instabilities}
\label{sec-inst}
%
%

Let us consider a small perturbation in the expanding system.
The amplitude of this perturbation may grow, decrease or
oscillate depending on the conditions.  In the presence of such a
perturbation the phase space distribution is:
$$
f(\tau,\vec{r},\vec{p}\ ) = f_0 + f_1 (\tau,\vec{r},\vec{p}\ ),
$$
with the normalization
$n  (\tau,\vec{r}\ )=\int d^3p\ (f_0+f_1)$ and
$n_1(\tau,\vec{r}\ )=\int d^3p\ f_1 $,
where the unperturbed density $n_0$ is homogeneous, $\vec{\nabla} n_0 = 0$,
and $f_1$ should be a local spherical perturbation which is a solution of
the Vlasov equation
\begin{equation}
\frac{\partial f_0}{\partial \tau} +
\frac{\partial f_1}{\partial \tau} +
\frac{ \vec{p} }{ m } \left[
           \frac{\partial f_0}{\partial \vec{r} } +
           \frac{\partial f_1}{\partial \vec{r} }
                                                  \right ] -
\vec{\nabla} U\
             \left[
             \frac{\partial f_0}{\partial \vec{p}}\ +
             \frac{\partial f_1}{\partial \vec{p}}\  \right] = 0 .
\label{e:vl3}
\end{equation}
The solution of the Vlasov equation is treated in details in
Ref.~\cite{kn:be88}. Here we separate the Vlasov
equation into two equations, one for $f_0$ and one for $f_1$.
Separating non-vanishing dominant zeroth order terms we get the equation
$$
\frac{\partial f_0}{\partial \tau} +
\frac{ \vec{p} }{ m } \frac{\partial f_0}{\partial \vec{r} } = 0 ,
$$
which is satisfied by $f_0$ as given in Eq. (\ref{e:f0}). The first order terms
yield the linearized equation
$$
\frac{\partial f_1}{\partial \tau} +
\frac{ \vec{p} }{ m } \frac{\partial f_1}{\partial \vec{r} } -
\vec{\nabla} U\
          \frac{\partial f_0}{\partial \vec{p}}\ = 0 .
$$
We intend to find perturbations which grow, leading to instabilities of the
system. Some modes of growing perturbations may arise in thermal
surrounding, and their rate is determined by thermal and viscous damping
\cite{vv94,ck92,ckz93,tl80,lt73}.
These are usually slower, and so other faster processes may come into play
also. Here we intend to study non-thermal, growing perturbations, which
may occur after the thermal freeze-out only, but they can be faster
than the thermally damped processes \cite{csm95,np90,ck92,ckz93,cc94,aram}.
The growth
rate of such perturbations is determined by the long range nuclear mean
field potential, $U(\vec{r})$.  Usually different non-thermal channels of
instability
open only when a given time is passed after the freeze-out at $\tau_{fr}$, and
we reach a threshold time, $\tau_{th}$.

There are two characteristic time scales in the system: (i) the longer
post freeze-out expansion between $\tau_{fr}$ and $\tau_{th}$, and (ii)
the rapid growth of instability which develops after $\tau_{th}$.
When studying the dynamics of rapidly growing instabilities we can
usually neglect the much slower dynamics of the post freeze-out
expansion.

Different configurations can and should be taken into account when studying
instabilities in a quenched (supercooled) system.

\subsection*{\large\sl 2.1. Plane wave perturbation}

The stability of the Vlasov equation against plane wave
perturbations was examined in detail by different groups
recently \cite{kn:rand,kn:mis}.
If we expand the perturbation, $f_1$, as
\begin{equation}
\label{e:p1}
f_1(\vec{r},\vec{p},t) = \sum_{k}\ f_{k}(p,t) {1 \over{
\sqrt{\Omega} }}\ e^{i\vec{k}\vec{r}} ,
\end{equation}
and search for the solution of $f_{k}(p,t)$ in the form
$$
f_{k}(p,t) = f_{{k,\omega}}(p)\ e^{i\omega t} \quad ,
$$
according to Ref. \cite{kn:rand} we get the dispersion relation
$\omega(k)$
\begin{equation}
\label{eq:om-rand}
1 = {\partial U(k) \over{ \partial n }} \int {d^3p \over{ (2\pi
\hbar)^3 }}\  {(\vec{k}\vec{p}\ )^2 \over{ (\vec{k}\vec{p}\ )^ - m^2\omega^2 }}
\ {\partial \tilde{f} \over{ \partial \epsilon }} \quad ,
\end{equation}
where $\tilde{f}$ is the static uniform solution, and $U(k)$ denotes
the Fourier component of the effective field $U(\vec{r}\ )$. The mode
corresponding to wavenumber $k$ will be
unstable, when the $\omega(k)$ frequency becomes imaginary. It was also
shown in Ref.~\cite{kn:rand}, that for zero temperature the condition of
instability can be written as
\begin{equation}
\label{e:inst}
{2\over 3} \epsilon_{F} + n {\partial U(k) \over{
\partial n }} < 0 \quad ,
\end{equation}
with $\epsilon_{F}$ Fermi kinetic energy. The expression
goes over for $k\to 0$ into the condition of
{\it mechanical instability} (i.e., the compressibility
becomes negative)
$$
\left( {\partial p \over{ \partial n }} \right)_{T=0} =
{\partial \over{ \partial n }} \left( n^2 {\partial \epsilon
\over{ \partial n }} \right) = {2\over 3} \epsilon_{F} + n
{\partial U(n) \over{ \partial n }} < 0 \quad ,
$$
where the potential $U(n)$ depends only on the homogeneous density.

In Ref.~\cite{kn:mis} the instability condition was examined for a one
dimensionally expanding ground state system. The acting force was a Skyrme
force with Yukawa surface term. The resulting instability condition reads as
Eq.~(\ref{e:inst}), where
$$
{\partial U(k) \over{ \partial n }} = -2 \beta
 + \gamma (\sigma + 1) (\sigma + 2) n_0^{\sigma} -
 {4 \pi V_0 \over{ \mu^2 + k^2 }}  \quad ,
$$
and $\beta$, $\gamma$, $\sigma$, $V_0$ and $\mu$ are the Skyrme and Yukawa
force parameters (see Eq.~(\ref{e:3p-1})).

In case of a uniformly expanding system the same condition
of instability for a plane wave perturbation can also be obtained
in a fashion similar to the spherical case discussed later,
see Appendix A. This approach leads to the following condition
\begin{equation}
1 = - \frac{2 \pi \tau_{fr} m C}{\tau}
\sqrt{2mT_{fr}}\ \frac{\partial U(k)}{\partial n}\
\int\limits_0^\infty
  \frac{d y\  \sqrt{y} }{ (y+s) [1+ e^{y-\mu_{fr}/T_{fr}} ] }\quad ,
\label{e:dis1}
\end{equation}
where $s= m \kappa^2 \tau^4 /( 2 k^2 T_{fr} \tau_{fr}^4) $ ,
$\kappa = i \omega$ and
$\partial U(k)/\partial n$ is the same expression as in
Ref.~\cite{kn:mis}.

\subsection*{\large\sl 2.2. Spherical drop/bubble perturbations}

In the following we want to study more realistic perturbations
instead of plane waves.
We consider local spherical bubbles, since we believe these are the
first instabilities which start to grow
\cite{pnb92}. In general spherical perturbations minimize the
surface and surface energy, so these can be formed the earliest.
(On the other hand plane wave perturbations may grow more rapidly
at later stages with stronger driving forces due to the increased
surface.)

Consider a spherical drop with a surface density profile
exponentially decreasing characterized by the parameter $k$,
a central density, $n_{\rm c}$, and radius $R(\tau)$
\begin{equation}
f_1 = f_s
\left\{
\begin{array}{ll}
n_{\rm c} \exp\left[ - k\frac{\tau_{fr}}{\tau}
         \left(r - R(\tau)\right) \right] \phantom{\frac{\tau_{fr}}{\tau}} &
      r \geq R(\tau)\\
      \phantom{-}& \\
n_{\rm c} & r <    R(\tau)
\end{array} \right. ,
\label{e:ps5}
\end{equation}
where
$
R(\tau) = R^*\ \frac{\tau}{\tau_{th}}\ e^{\kappa (\tau-\tau_{th})}
$
is the $\tau$ dependence of the radius after the threshold time,
$\tau_{th}$ when the droplet becomes bigger than the critical radius
and it will be able to grow.
We are interested in the initial growth rate of the radius only, so that
in $R(\tau)$ the exponential term can be expanded into a power series for
small $\tau-\tau_{th}$, i.e.,
$$
R(\tau) \approx  R^*\ \frac{\tau}{\tau_{th}}\ [1+ \kappa (\tau-\tau_{th})] .
$$
Inserting this approximation into Eq. (\ref{e:ps5}) we get for the
exterior part ($r > R(\tau)$) of the profile:
$$
f_1 = f_s  n_{\rm c}
    \exp\left(   k R^* \frac{\tau_{fr}}{\tau_{th}}  \right)\
    \exp\left( - k \frac{\tau_{fr}}{\tau} r +
      k R^* \frac{\tau_{fr}}{\tau_{th}} \kappa (\tau-\tau_{th}) \right) .
$$

Since  we  consider small perturbations only, where the linearization of
Eq.(~\ref{e:vl}) holds, this solution is valid only
for a short time after the instability starts to grow.
The dispersion relation will lead to a dynamical growth factor,
$\kappa$, depending both on $k$ and $R$.

(In the plane wave expansion of the perturbation studied in
Ref.~\cite{kn:rand,kn:mis} the opening of the channel of the instability was
indicated when $\omega$ became imaginary.
Thus perturbations preceding $\tau_{th}$ did not grow.  Our approach is
basically equivalent to their one presented here, however, we wanted to
emphasize that the instability may grow only after $\tau_{th}$.)

Since $f_s(\tau,\vec{P})$ has a characteristic time dependence
on the slow scale (i) of the post freeze-out expansion, we assume that its
time-derivatives are negligible compared to other time-derivatives of
$f_1(\tau, \vec{r}, \vec{p}\ )$ corresponding to rapid inequilibrium processes
(ii) like\ \  $\exp[\kappa(\tau - \tau_{th})]$.  Furthermore, we assume that
$f_s$ depends on the LR momentum, $\vec{P}$ only, i.e., it does not have any
other dependence on $\vec{r}$ other than what is included in $\vec{P}$.

We search for such solutions of the Vlasov equation,
$f_1(\tau,\vec{r},\vec{p}\hspace*{2pt})$ and $f_s(\tau,\vec{P})$. We assume
that
such solutions can be obtained only some time, $\tau$, after the  freeze-out
at $\tau_{fr}$, i.e. at $\tau_{th} > \tau_{fr}$. Before this time thermal
processes and thermal damping is dominant which generally lead to
slower nucleation than post-freeze-out processes driven by the
background fields.

We can calculate the critical droplet radius, $R^*_{\rm crit}$. Droplets
smaller than $R^*_{\rm crit}$ tend to disappear, while droplets larger than
$R^*_{\rm crit}$ may
start to grow.  Thus we will study the growth rate of critical size droplets.
The critical radius, $R^*_{\rm crit}$ is calculated in Appendix E.

The critical radius, $R^*_{\rm crit}$, should be evaluated when the channel of
instability opens at $\tau_{th}$, and the critical droplet just starts to
grow. In the 3-dimensional scaling expansion the critical radius scales with
the overall scaling, which leads to a quasi-static critical radius of
$R^*_{\rm crit} \tau/\tau_{th}$.

The critical radius, $R^*_{\rm crit}$, depends on the background
nucleon density at the
opening of the instability, $n_0(\tau_{th}) = n_{fr} (\tau_{fr}/\tau_{th})^3$,
or consequently on $\tau_{th}$, furthermore on the surface parameter, $k$,
on the central density, $n_{\rm c}$, and on the parameters of the interaction
potential. The total energy of the system can be simultaneously minimized
by varying $k$ and $n_{\rm c}$, and searching for an extremum as a function of
$R^*$.

As we mentioned the time-scale of the expansion is assumed to be slow
compared to the
time-scale of the instability, so $R^*$ can be considered as a time
independent constant when studying the growth of instability.

The perturbation (\ref{e:ps5}) satisfies the Vlasov equation both
for the exterior and interior region, if an averaged Yukawa potential is used
(see Appendix C).

The dispersion relation for such a spherical perturbation can be written as
\begin{equation}
1 = - \frac{2 \pi \tau_{fr} m C}{\tau} \sqrt{2mT_{fr}}\
\frac{\partial U}{\partial n}\
\int\limits_0^\infty
  \frac{d y\  \sqrt{y} }{ (y-S) [1+ e^{y-\mu_{fr}/T_{fr}} ] }
\label{e:dis3}
\end{equation}
(see Appendix B), where $S= m \kappa^2 (R^*)^2 \tau^4
/( 2 T_{fr} \tau_{fr}^2 \tau_{th}^2) $ .

It is easy to see, that for $S=0$ without
the surface term this condition is equivalent to the isothermal mechanical
instability even for $T\neq 0$ (Appendix D), that is the region of dynamical
instability and of mechanical one coincide.
Although the direct $k$-dependence drops out of the dispersion
relation, but since $R^*$ depends on $k$ so the dispersion relation
is still applicable.  It is interesting to mention that both
dispersion relations, (\ref{e:dis1},\ref{e:dis3}), yield
the same condition for evaluating the threshold time for the perturbation
which is just on the boundary to be able to grow, i.e., $s=S=0$.

The features of the potential and the Yukawa term in it are vital
in determining the properties of the static, critical droplet or bubble.
Physically we can consider two situations in the
course of final multifragmentation.

Depending on the beam energy, after the initial compression we reach
the most compressed and heated up state with a definite specific entropy.
This stage is then followed by an expansion, which is adiabatic
to a good approximation.
If the final specific entropy is smaller than the critical entropy of the
nuclear liquid-gas phase transition, the expansion will lead to a stretched
(or quenched) liquid state, with density $n_0$ below the
normal nuclear density, $n_N$.  The instabilities will lead then to bubble
formation with an interior nuclear gas phase density, $n_1 + n_0 \approx
0.1-0.4 n_N$.
If on the other hand the final specific entropy exceeds the critical entropy,
the expansion will lead to a oversaturated (or quenched) nuclear vapor state,
with density $0.1-0.4 n_0$, below the critical nuclear density.  The
instabilities will lead now to the condensation of a nuclear liquid droplet
with an interior nuclear density, $n_1 + n_0 \approx n_N$.
\bigskip

\section*{\large\bf 3. Condition for the instabilities to grow}
\label{sec-cond}

Equations~(\ref{e:dis1},\ref{e:dis3}) determine
the condition for $\kappa$ becoming real, but its sign remains undefined. From
equations~(\ref{e:p1},\ref{e:ps5}) it is easy to see, that the
amplitude of the perturbation will depend on the sign of $\kappa$: positive
$\kappa$-s will cause exponentially increasing perturbation, while
perturbations corresponding to negative $\kappa$ will be damped rapidly. To
determine the sign of the $\kappa$ we have to see, how the energy changes due
to the perturbation.  If the configuration with the perturbation acquires
smaller total nuclear energy (the total energy without the flow) than the
unperturbed system can we speak about
growing
instabilities, that is flow can develop to take extra matter into the
perturbation.

In the following we consider density dependent Skyrme forces with an averaged
Yukawa term.
The total nuclear energy of the system can be written as
$$
E =   E_{kin}
    - \beta \int d^3r\ n^2(\vec{r}\ )
    + \gamma \int d^3r\ n^{\sigma+2}(\vec{r}\ ) -
\hspace*{1.5cm}
$$
\begin{equation}
V_0 \int d^3r_1 d^3r_2 \ n(\vec{r}_1) n(\vec{r}_2)
{e^{-\mu \mid \vec{r}_1 - \vec{r}_2 \mid}
\over{\mid \vec{r}_1 - \vec{r}_2 \mid} } ,
\label{e:3p-1}
\end{equation}
where the values of $\beta$, $\gamma$, $\sigma$, $V_0$ and $\mu$ are the same
as in Ref.~\cite{kn:mis} and summarized in Table 1.
\begin{table}[t]
\begin{center}
\begin{tabular}{||c|c|c|c|c|c||} \hline \hline \\[-4mm]
 & $\beta$ (MeV fm$^3$) & $\gamma$ (MeV fm$^{3 ( \sigma + 1 )}$) & $\sigma$ &
 $V_0$ (MeV fm) & $\mu$ (fm$^{-1}$) \\ \hline \hline
SOFT & 1051.76 & 1107.41 & 1/6 & 83.5 & 2 \\ \hline
HARD & 365.0 & 808.65 & 1 & 83.5 & 2 \\ \hline \hline
\end{tabular}
\end{center}
\caption{The parameterization of the SOFT and HARD EOS}
\end{table}
Let us consider a system which is initially
homogeneous and has constant density $n_0(\tau)=n_{fr} (\tau_{fr}/\tau)^3$.
Introducing now a small perturbation in the density in a way that the total
mass number has to be conserved, we obtain a density distribution
\begin{equation}
n(\tau,\vec{r}\ ) = n_0(\tau) + n_1(\tau,\vec{r}\ ) - {\Gamma \over{ \Omega }}
\label{e:3p-2}
\end{equation}
where $\Gamma=\int d^3r\ n_1(\tau,\vec{r}\hspace*{2pt})$, $\Omega$ is the
volume of the
system, and $n_1$ is assumed to be small compared to $n_0$.
If the initial configuration is such that the formation of a perturbation
may lead to a decrease of the energy such perturbations will appear
and grow spontaneously. This will lead to a multifragmentation of the system.
We consider here the energy of the system and not the Helmholtz free energy
because we are describing a post freeze-out situation when we do not have a
heath bath any more.

Substituting (\ref{e:3p-2}) into (\ref{e:3p-1}) and expanding in terms of
$n_1$ up to the second order, we evaluate the total energy of the perturbed
system.  For large enough systems the terms containing $\Gamma^2/\Omega$ can
also be neglected.  Thus the total nuclear energy can be written
as $E=E_0+\Delta E$,
where
\begin{equation}
{E_0\over A} = {E_{\rm {kin}}(n_0) \over A} - \left( \beta +
{4\pi V_0 \over{ \mu^2 }} \right) n_0 + \gamma n_0^2 \quad ,
\label{e:3p-3}
\end{equation}
and
$
A = \int d^3r\ n(\tau,\vec{r}) = \int d^3r\ n_0 = n_0 \Omega .
$
The change of the energy due to the formation of the perturbation
in the second order of $n_1$ (the first order terms cancel due to the mass
number conservation: Eq.~\ref{e:3p-2})
is then
\begin{eqnarray}
\Delta E
&=&
 \Delta E_{\rm {kin}}
- \left( \beta
 + {4\pi V_0 \over{\mu^2 }}
 - {1\over 2} (\sigma + 1) (\sigma + 2) \gamma n_0^{\sigma}
  \right)  \overline{N}_1
+ \Delta \varepsilon_{\rm {surf}} \overline{N}_1 ,
\nonumber \\
\Delta \varepsilon_{\rm {surf}}
&=&
  \left[ {4\pi V_0 \over{ \mu^2 }} -
 {V_0 \over{ \overline{N}_1 }}
 \int d^3r_1 d^3r_2\ n_1(\vec{r}_1) n_1(\vec{r}_2)
 {e^{-\mu \mid \vec{r}_1 - \vec{r}_2 \mid}
 \over{\mid \vec{r}_1 - \vec{r}_2 \mid} }
\right] ,
\label{e:3p-4}
\end{eqnarray}
where
$ \overline{N}_1 = \int d^3r\ n_1^2(\vec{r}) $.
The term $\Delta \varepsilon_{\rm {surf}}$ is the surface correction due to
the Yukawa interaction.
The sign of $\Delta E$ will determine whether an instability
may increase or will be damped.

It is not immediately clear, whether the kinetic energy gives a contribution
to $\Delta E$ or not. In thermal systems at high temperature and low
density, where the exact value of the Fermi momentum is not too important we
can assume that $f = f_0 + f_1$, where $f_1 \approx [n_1(\tau,\vec{r})/n_0]\
f_0$, and the kinetic energy depends only linearly on $n_1$, that is a
density perturbation will not cause a change of total kinetic energy.  For a
isotherm, degenerate system, where the kinetic energy is nearly proportional
to $n^{5/3}$, a perturbation may give a contribution to $\Delta E$.  Here we
are studying a post freeze-out situation out of thermal equilibrium.  Now the
kinetic energy depends on the form of our perturbed phase space distribution
function, $f_s(\tau,\vec{P})$, we choose or obtain.  This may have different
characteristics depending on the density of our frozen-out system. Thus we
will examine both situations, the one without the kinetic energy contribution
($\Delta E_1$), and the one with ($\Delta E_2$).

\begin{figure}
\vspace*{-10mm}
       \begin{center}
       \leavevmode
       \epsfxsize=12.6cm\epsfbox{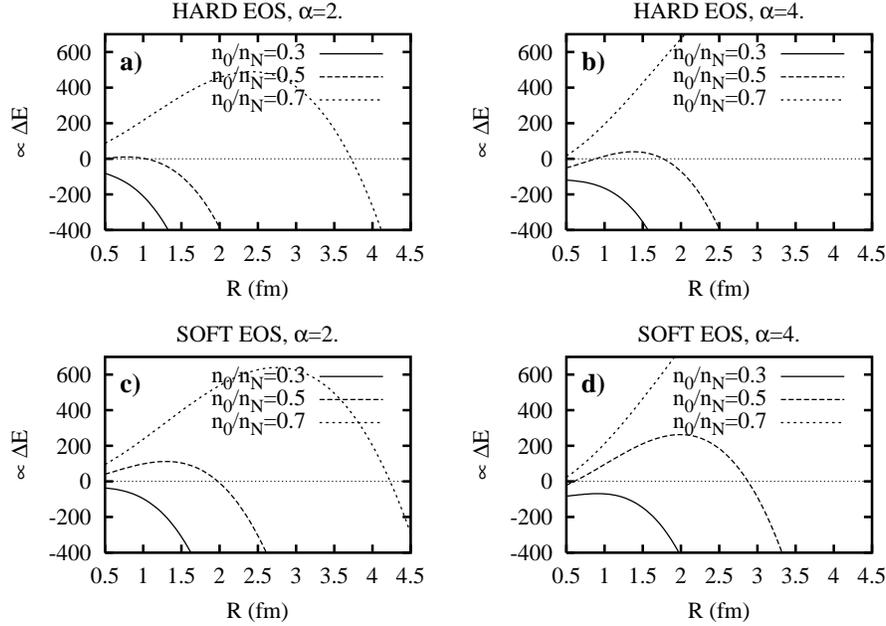}
       \end{center}
\vspace*{-10mm}
\caption{\small The dependence of the energy difference, $\Delta E$,
caused by the formation of a bubble of radius $R$, without the kinetic term
on the radius of the bubble
for two different diffuseness coefficients, $\alpha=
k \tau_{fr}/\tau$, and two equations of state.
The energy difference is evaluated for three post-freeze-out densities
$n_0/n_N = $ 0.3, 0.4 and 0.7.  The maxima of the curves (if any)  is at the
critical radius, $R^*$
}
\end{figure}
The surface correction, $\Delta \varepsilon_{surf}$, for the cases of
plane wave and spherical perturbations considered in Section 2 are:
\begin{eqnarray}
{\rm {Case\ A}}\ \quad & {4\pi V_0 \over { \mu^2 }} \left( 1 - {\mu^2 \over{
\mu^2 + k^2 }}  \right)
\nonumber \\
{\rm {Case\ B}}\ \quad & ({4\pi V_0 /{ \mu^2 }})
\left[ 1 - g(R^*,k) \right]
\nonumber
\end{eqnarray}
where $g$ is a complicated function of $R^*$ and $k$ (see
Appendix C). For the used forces in Eq.~(\ref{e:3p-1}) in the case A
$\Delta E$ turns out to be as seen in Ref.~\cite{kn:mis}.
$$
\Delta E_{kin} - \left( \beta + {4 \pi V_0 \over{ k^2 + \mu^2 }}
  - {(\sigma+1) (\sigma+2) \over 2 } \gamma n_0^{\sigma} \right)
  \overline{N}_1 < 0 .
$$

We get a negative change in the energy only for $k < k_{\rm crit}$, where
the critical value of $k$ is
$$
k^2_{\rm crit} = -\mu^2 + {4 \pi V_0 \over{ {\Delta E_{kin} \over{
\overline{N}_1 }}
- \beta + {(\sigma+1) (\sigma+2) \over 2 } \gamma n_0^{\sigma} }}
$$
In case B
the required negativity of $\Delta E$ can be expressed in a more complicated
way.

\begin{figure}[t]
\vspace*{-10mm}
       \begin{center}
       \leavevmode
       \epsfxsize=12.6cm\epsfbox{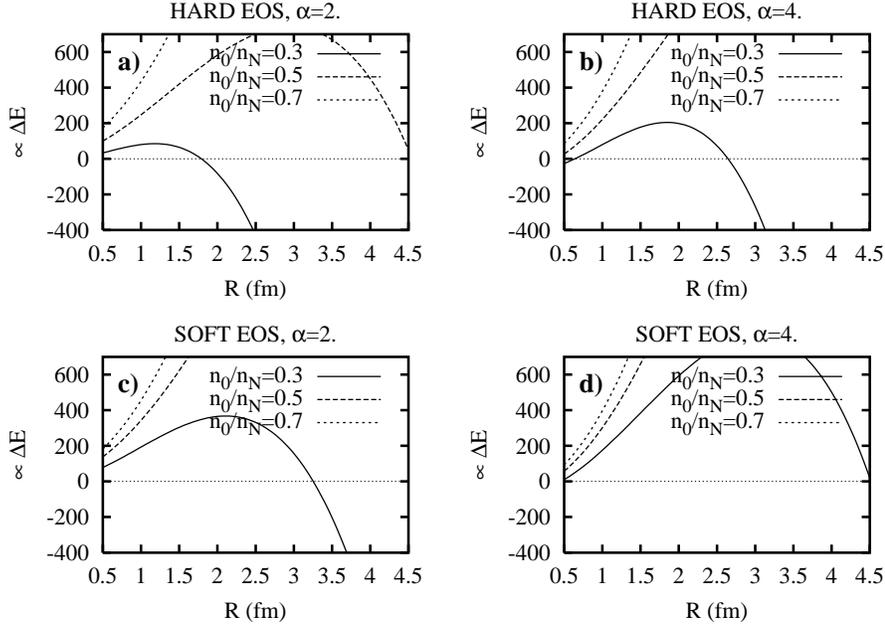}
       \end{center}
\vspace*{-10mm}
\caption{\small The same as Fig. 1 with kinetic energy term included}
\end{figure}
\section*{\large\bf 4. Results}

In the following we present here the results for a spherical droplet
perturbation considered in Section 2.2. The total energy of the system is given
in Eq.~(\ref{e:3p-1}) and the energy change due to the bubble formation in
Eq.~(\ref{e:3p-4}). The surface energy,
$\Delta\varepsilon_{\rm surf}$,
energy given in
Eq.~(\ref{e:as1}) depends strongly on $\alpha = k \tau_{fr}/\tau$ and on the
radius $R^*$ of the bubble. To see the effect of the bubble shape on the energy
change $\Delta E$ we give this change for different $\alpha$ and
density values as the function of the bubble radius.
In Fig.~1 we
assumed that the kinetic energy does not give any contribution in second order
to the energy, in Fig.~2 we considered a ground state Fermi kinetic energy
contribution. As one can see, the effect of the surface term is larger if we
have sharper surfaces (larger $\alpha$ values). As one expects, the effect of
the surface energy,
$\Delta\varepsilon_{\rm surf}$,
decreases for large radii.
As a first step we considered the solution of Eq.~(\ref{e:dis3}) for
$\kappa$=0, and compared the condition of the dynamical instabilities to grow
with that of the mechanical instability for infinite systems.
Without surface term the two condition are the same,
as was pointed out already in Ref.~\cite{kn:rand}. With the surface term
the instability region decreases, just as in Ref.~\cite{kn:mis}. In Fig.~3 we
give the $n$ -- $T$ curve of instability region for different $\alpha$ and
$R^*$ values.
With small supercooling first big droplets can nucleate and grow, then with
stronger supercooling smaller droplets can also be formed.
For big droplets ($R^*$ = 4.5 fm) the effect of the surface term
is almost negligible.
If we wait longer after the freeze-out in the expanding system, i.e., we
increase $\tau$ and thus having a smaller $\alpha$, $(\alpha=\kappa
\tau_{fr}/\tau)$ we have the possibility of instabilities earlier, with
smaller supercooling.
The calculations are done both for soft and hard equation of state.
One sees, that the effect of the surface term is more significant for the soft
equation of state.

As the next step we want to determine the break up of the system starting from
different freeze out densities times and temperatures. A reasonable freeze out
density should be in the range of $n_0=$0.08 -- 0.14 fm$^{-3}$.
In our calculation we
choose $n_{fr}$ = 0.1 fm$^{-3}$. Other freeze out densities can be considered
simply rescaling $\tau_{fr}$, $\mu_{fr}$ and $T_{fr}$ to keep the relation
$(n_0/n_{fr})^{2/3} = (T/T_{fr}) = (\tau_{fr}/\tau)^2$ and
$\mu_{fr}/T_{fr}$ constant (see the remarks after Eq.~(\ref{e:f0})).
The freeze out time
$\tau_{fr}$ defines the flow energy of the system as ${\displaystyle
E_{\rm flow}/A = \frac{3}{5} {R^2 \over{ \tau_{fr}^2 }} }$
for a system with radius $R$. We
assume, that the total excitation energy of the system is large enough to reach
the break up densities \cite{pw95}, so if that condition is fulfilled, the
freeze out time
is not defined in the model, and the time scale is not fixed.
We followed the paths along the trajectories in the $n$ -- $T$ plane
from different initial freeze out configurations.
The instability condition Eq.~(\ref{e:dis3}) is examined along the
trajectories, and the time of the solution for growing instability (break up)
can be expressed as ${\displaystyle \Delta \tau = \tau - \tau_{fr} =
\tau_{fr} \left[ \left( {n_{fr} \over n} \right)^{1/3} - 1 \right] }$
after the freeze out.
\begin{figure}[t]
\vspace*{-10mm}
       \begin{center}
       \leavevmode
       \epsfxsize=12.6cm\epsfbox{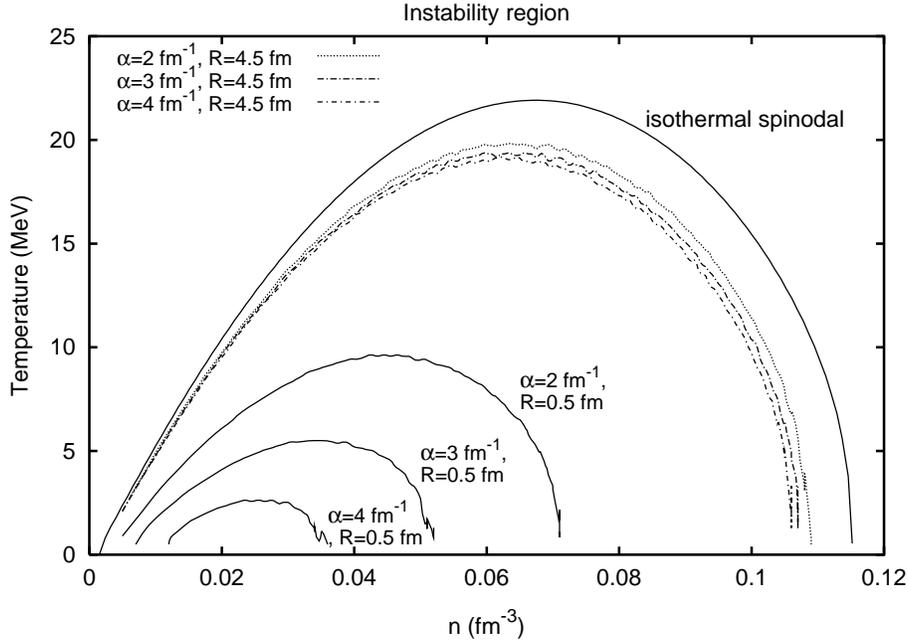}
       \end{center}
\vspace*{-10mm}
\caption{\small The dynamical and mechanical instability regions.
The region of mechanical instability is bordered by the isothermal
spinodal, where the isotherm sound speed vanishes and becomes imaginary.
For large size (radius $R$) of the critical bubbles or for systems without
surface energy the dynamical instability region is the same as the mechanical
one.
The break up instability depends on the radius $R$
and on the stretching of the system parameterized by $\alpha$}
\end{figure}

In Fig.~4 we show the part of the trajectories
of the post-freeze-out expansion denoted by dotted line where
the instability condition is fulfilled and the energy change, $\Delta E_1$,
is negative. We found, that whenever condition~(\ref{e:dis3}) is
fulfilled, this energy change is always negative. The more strict condition,
using
$\Delta E_2$, however, excludes some part of the trajectories (solid line) at
temperatures above 10 MeV for the hard equation of state. For the soft equation
of state there is no such exclusion, nevertheless,
the instability region is smaller.
We give the results for the two equations of state and different parameters
for both.
For comparison we show the boundary of the isothermal mechanical instability
(which corresponds to the $R^* \to \infty$ situation). In the parameter region
we count as physical ($R^* \approx $ 2--3 fm, $k \approx $ 4 -- 6 fm$^{-1}$)
there are no significant changes on this parameters.
\begin{figure}
\vspace*{-10mm}
       \begin{center}
       \leavevmode
       \epsfxsize=12.6cm\epsfbox{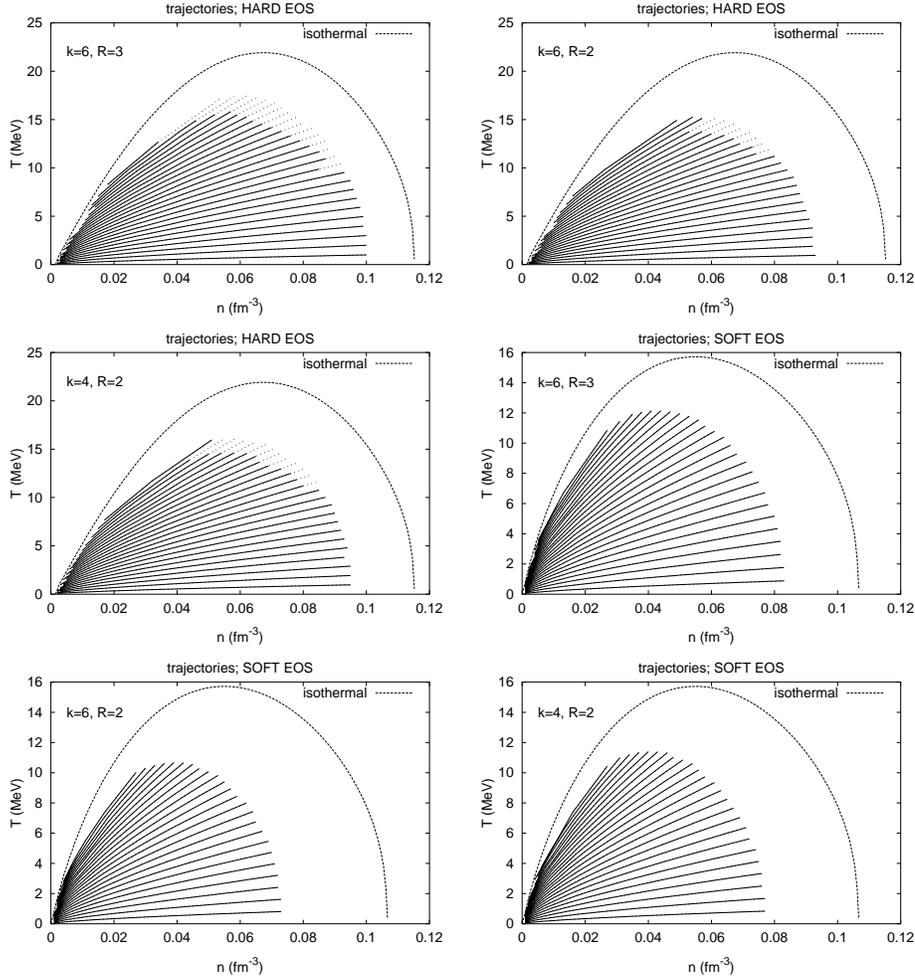}
       \end{center}
\vspace*{-10mm}
\caption{\small The trajectories of an adiabatic
post-freeze-out expansion
starting at the same, $n_{fr}=0.1$ fm$^{-3}$ freeze out
density. The different lines are originated from different freeze out
temperatures. The solid lines
correspond to the case where the energy difference with the kinetic term is
negative, the dotted line is the case where the energy difference without the
kinetic term is negative. The latter ones define the wider region.
The possibility of  dynamical post-freeze-out instability opens only
after some penetration into the domain of the isothermal spinodal,
while thermally dominated homogeneous nucleation may start immediately,
although slowly
}
\end{figure}

Following the trajectory of
the post-freeze-out expansion starting from a given initial
$n_{fr},T_{fr}$ configuration the
instability condition Eq.~(\ref{e:dis3}) has solutions for different
$S$ or $\kappa$ values. For high densities and temperatures $S$ is negative,
that is there are no real solution for $\kappa$.
As the density (and correspondingly the temperature) decreases it
continuously becomes positive. The speed of growth of the instabilities is
determined by ${\displaystyle \kappa = {\tau_{fr} \over{ \tau }} {\tau_{th}
\over{ \tau }}  \sqrt{ {2 T_{fr} S \over{ m (R^*)^2 }} } }$. Evaluating this
expression we assumed that $\tau_{th} / \tau \approx 1$. In Fig.~5
$\kappa$ is shown along given expansion trajectories
($k$=4 fm$^{-1}$ $R$= 2 fm,
$T_{fr}$ = 1 MeV, 16 MeV and 21 MeV) for the hard equation of state.
The speed $\kappa$, the instabilities are developing with, is changing,
first it increases as the nucleus
evolves to smaller densities, later it decreases back to zero.
The time scale of the expansion for the radius of the perturbations
 from Fig.~5 gives $\approx$ 20 fm/c for the
$\kappa \approx $ 0.06. However, the growth rate of the instability from
Eq.~(\ref{e:ps5}) is a double exponential: $n_c\ exp\left( k \tau_{fr} /
\tau_{th}\ R^* e^{\kappa \Delta (\tau-\tau_{th})} \right)$, which is much
faster.
This region of the exponentially developing instabilities breaks up the
system.
\begin{figure}[t]
\vspace*{-10mm}
       \begin{center}
       \leavevmode
       \epsfxsize=12.6cm\epsfbox{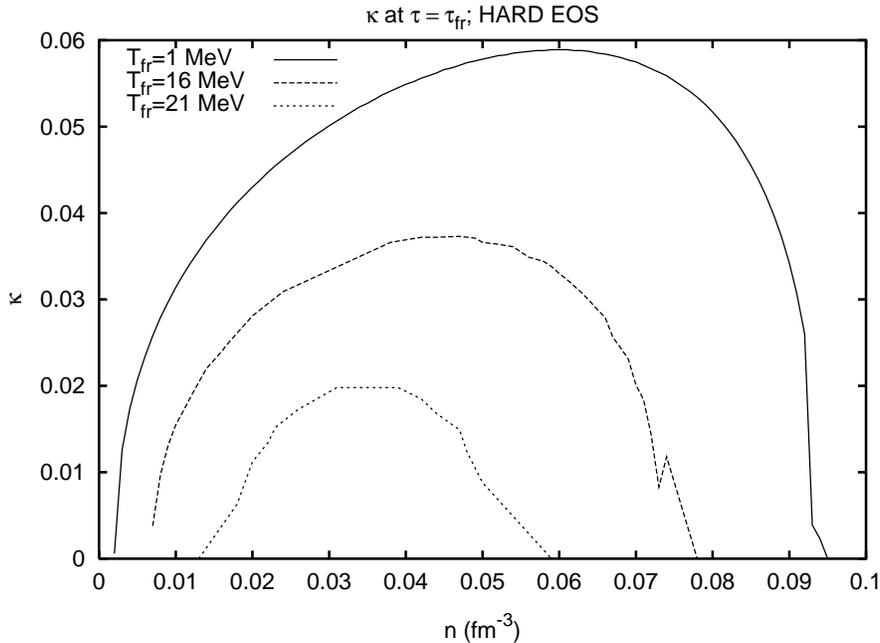}
       \end{center}
\vspace*{-10mm}
\caption{\small The growth parameter, $\kappa$, along trajectories
of post-freeze-out expansion starting at $n_{fr}=$ 0.1 fm$^{-3}$ and
at the given freeze-out temperatures
}
\end{figure}

\section*{\large\bf Appendix A:
Dispersion relation for plane wave perturbation}
\label{sec-plane}

Let us consider a plane wave perturbation in another form than in
Ref. \cite{kn:rand} emphasizing our time scale
\begin{equation}
f_1(\tau, \vec{r}, \vec{p}) =
         f_s(\tau, \vec{P})\
{\exp \left[ i \frac{\vec{k}\tau_{fr}}{\tau} \vec{r} +
\kappa (\tau-\tau_{th}) \right]} ,
\label{e:pp1}
\end{equation}
where it is taken into account, that the wave number, $\vec{k}$,
scales with $\tau$, as all other parameters of the flow. Thus
$k$ is a constant, independent of $\tau$.
Here  $f_1$ is a phase space distribution  which should
satisfy the Vlasov equation (\ref{e:vl}), and it represents a plane wave
with growing amplitude if $\kappa$ is a real positive number.
Since  we  assume small perturbations only, this solution is valid only
for a short time after the instability starts to grow.
(Frequently in similar studies the perturbation is studied in a form
containing an $\exp(i\omega  \tau)$ term, and the opening of
the channel of instability is indicated when $\omega$ becomes imaginary.
Thus perturbations preceding $\tau_{th}$ will not grow.  This approach is
basically equivalent to the one presented here, however, we wanted to
emphasize that the instability may grow only after $\tau_{th}$.)

As it was already mentioned we also assume  that $f_s(\tau,\vec{P})$
has a characteristic time dependence
on the slow scale (i) of the post freeze-out expansion, so that its
time-derivatives are negligible compared to other time-derivatives of
$f_1(\tau, \vec{r}, \vec{p})$ corresponding to rapid inequilibrium processes
(ii) like $\exp[\kappa(\tau - \tau_{th})]$.  Furthermore, we assume that
$f_s$ depends on the LR momentum, $\vec{P}$ only, i.e., it does not have any
other dependence on $\vec{r}$ other than what is included in $\vec{P}$.

We search for such solutions of the Vlasov equation,
$f_1(\tau,\vec{r},\vec{p})$ and $f_s(\tau,\vec{P})$. We assume that
such solutions can be obtained only some time, $\tau$, after the  freeze-out
at $\tau_{fr}$, i.e. at $\tau_{th} > \tau_{fr}$.

Using the above form, (\ref{e:p1}), of perturbation the density change
arising from this perturbation is:
$$
n_1(\tau,\vec{r}) = n_s(\tau)\ \exp[i \frac{\vec{k}\tau_{fr}}{\tau}
\vec{r} + \kappa (\tau-\tau_{th})] .
$$
Inserting a plane wave perturbation
Eq. (\ref{e:p1})
into the Vlasov equation we obtain for
the force used in Eq.~(\ref{e:3p-1})
$$
f_s(\tau,\vec{P})\ \left[\kappa +
  \frac{ i \vec{k}\tau_{fr} }{ m \tau }
  \left( \vec{p} -  m \vec{r} / \tau \right) \right]
- \hspace*{1.5cm}
$$
\begin{equation}
\frac{\partial U(k)}{\partial n}\
n_s(\tau)  \frac{i \vec{k}\tau_{fr}}{\tau} \
\left( - \frac{f_0^2}{C} \right)
\frac{[\vec{p} - m \vec{r}/\tau]}{mT_{eff}}
      \exp\left[\frac{[\vec{p} - m \vec{r}/\tau]^2}{2mT_{eff}} -
       \frac{\mu_{eff}}{T_{eff}}\right]
= 0 \quad  ,
\label{e:vls}
\end{equation}
where
$$
\frac{\partial U(k)}{\partial n}\ =  -2 \beta
 + \gamma (\sigma + 1) (\sigma + 2) \ n_0^{\sigma}
  - {4 \pi V_0 \over{ \mu^2 + k^2 }}
$$
as in Ref.~\cite{kn:mis}.

Note that the momentum appears only inside expressions of the
LR momentum,
$\vec{p} - m \vec{r}/\tau$, thus we can integrate it out in the LR frame.
Thus from (\ref{e:vls}) we can express $f_s$, and integrating it
over the LR momentum, $\vec{P} = \vec{p} - m \vec{r} / \tau$,
we obtain $n_s(\tau)$, which then can be eliminated from
both sides yielding:
\begin{equation}
1 = - \frac{1}{T_{eff} C}\ \frac{\partial U(k)}{\partial n}\
\int d^3 P   f_0^2(\vec{P})
       \frac{i \tau_{fr} \vec{k} \vec{P} / \tau }
{ m \kappa + \frac{i \vec{k}\tau_{fr}}{\tau} \vec{P} }
       \exp\left[\frac{\vec{P}^2}{2mT_{eff}} -
       \frac{\mu_{eff}}{T_{eff}}\right]
\label{e:vls2}
\end{equation}
We can separate the variable of the integral to a parallel, $P_{||}$, and
an orthogonal, $\vec{P}_\perp$, component with respect to $\vec{k}$,
and the integration over the components perpendicular to $k$ can be
performed. This yields
\begin{equation}
1 = - 2\pi m C\ \frac{\partial U(k)}{\partial n}
\hspace*{-1mm}
\int\limits_{-\infty}^\infty \hspace*{-1mm} d P_{||}
  \frac{i k P_{||}\tau_{fr}/\tau}{m \kappa + i k P_{||} \tau_{fr} / \tau }
  \left\{ 1+ \exp\left[\frac{P_{||}^2}{2mT_{eff}} -
       \frac{\mu_{eff}}{T_{eff}}\right]  \right\}^{-1} \hspace*{-4mm} .
\label{e:vls3}
\end{equation}
Only symmetric functions contribute to this integral, so we symmetrize it by
multiplying both the numerator and denominator by
$m \kappa - i k P_{||}\tau_{fr} / \tau $,
and then dropping the antisymmetric term
we end up having
\begin{equation}
1 = - 4\pi m C\ \frac{\partial U}{\partial n}
\hspace*{-1mm}
\int\limits_0^\infty \hspace*{-1mm} d P_{||}
  \frac{P_{||}^2 }{ [ m \kappa \tau / (k\tau_{fr}) ]^2 + P_{||}^2 }
  \left\{ 1+ \exp\left[\frac{P_{||}^2}{2mT_{eff}} -
       \frac{\mu_{eff}}{T_{eff}}\right]  \right\}^{-1} \hspace*{-4mm} .
\label{e:vls4}
\end{equation}
Introducing a new variable, $y = P_{||}^2 / (2 m T_{eff})$,
a straightforward calculation will lead to the dispersion relation in the
integral form:
\begin{equation}
1 = - 2 \pi m C \sqrt{2mT_{eff}} \ \frac{\partial U}{\partial n}\
\int\limits_0^\infty
  \frac{d y\  \sqrt{y} }{ (y+s) [1+ e^{y-\mu_{fr}/T_{fr}} ] },
\label{e:adis1}
\end{equation}
where $s= m \kappa^2 \tau^2 /( 2 k^2 \tau_{fr}^2 T_{eff}) =
          m \kappa^2 \tau^4 /( 2 k^2 T_{fr} \tau_{fr}^4) $ .

\section*{\large\bf Appendix B: Dispersion relation for spherical
perturbation}
\label{sec-yukpot}

For the sake of simplicity let us first study a spherical cusp
perturbation
centered around some
interior point $\vec{r}_c(\tau) = \vec{r}_0 \tau / \tau_{fr}$ of the type
\begin{equation}
f_1 = \exp\left[ - k\tau_{fr} |\vec{r} - \vec{r}_0 \tau / \tau_{fr} |/\tau
+ \kappa (\tau-\tau_{th}) \right]
f_s ,
\label{e:ps0}
\end{equation}
where the center of the perturbation moves along the scaling expansion, and
this
center was at point $\vec{r}_0$   at the time of the freeze-out.
As we discussed it in the introduction we can assume that $\vec{r}_0 = 0$,
so without loosing the generality of the assumption, $|\vec{r}-\vec{r}_0|
\longrightarrow |\vec{r}| = r $,
 since the interior
points of the expanding nuclear system are equivalent. Thus
\begin{equation}
f_1 = \exp\left[ - k\tau_{fr} r/\tau +
\kappa (\tau-\tau_{th}) \right] f_s ,
\label{e:aps1}
\end{equation}
can serve to study the perturbation just as well.
Although this functional form of perturbation has a singularity in the center
this will not be essential for our study, and it could be removed by assuming
more complicated functional forms for the perturbation which would not show
such a singularity.

However, including the center of the perturbation, $\vec{r}_c$, explicitly
will allow us later to discuss interactions (e.g. fusion, repulsion, etc.)
of two (or more) elementary perturbations. Thus we will follow this somewhat
more complicated derivation although it is not necessary at the moment.

We will not explicitly define the form of $f_s$ at this stage, unlike in the
case of plane wave perturbations. Instead we will consider the integrals of
$f_1$ and $f_s$.
First the norms:
$$
n_1 = \int d^3p\  f_1 = \exp\left[ - k\frac{\tau_{fr}}{\tau} \left| \vec{r} -
\vec{r}_0 \frac{\tau}{\tau_{fr}}\right| + \kappa (\tau-\tau_{th})
\right] n_s ,
$$
and $n_s = \int d^3p f_s $, where we require that $n_s = n_s(\tau)$ does not
depend on $\vec{r}$.  Second the projections orthogonal to $(\vec{r} -
\vec{r}_c)$, i.e.,
$$
g_1(\tau, \vec{r}, P_{||}) =
2\pi \int\limits_0^\infty P_\perp d P_\perp f_1 =
\exp\left[ - k\frac{\tau_{fr}}{\tau}
\left|\vec{r} -
\vec{r}_0 \frac{\tau}{\tau_{fr}} \right| +\kappa (\tau-\tau_{th}) \right] g_s .
$$
Here $g_s(\tau, P_{||}) = 2\pi \int P_\perp d P_\perp f_s $, where we require
that $g_s = g_s(\tau, P_{||})$ does not depend on $\vec{r}$. We chose our
coordinate system so that $P_{||}$ is parallel to $(\vec{r} - \vec{r}_c)$,
and $\vec{P}_\perp$ is orthogonal to it.  These constraints are the required
implicit constraints on the choice of $f_s$.

Inserting Eq. (\ref{e:ps0}) into the Vlasov equation and integrating it
over $d^2 P_\perp$ we obtain
$$
g_s\
\left[\kappa
+
\frac{ k\tau_{fr} }{ m \tau }  \vec{p}
\frac{\vec{r} - \vec{r}_0 \tau/\tau_{fr}}{|\vec{r} - \vec{r}_0 \tau/\tau_{fr}|}
-
\frac{ k\tau_{fr} }{ \tau^2 }
 |\vec{r} - \vec{r}_0\tau/\tau_{fr}|
-
 \frac{ k \vec{r}_0 }{ \tau }
 \frac{\vec{r} - \vec{r}_0 \tau/\tau_{fr}}{|\vec{r} -
 \vec{r}_0 \tau/\tau_{fr}|}
\right] -
$$
\begin{equation}
2\pi  C n_s \frac{\partial U}{\partial n}\ \frac{k\tau_{fr}}{\tau} \
\frac{\vec{r} - \vec{r}_0 \tau/\tau_{fr}}{|\vec{r} - \vec{r}_0 \tau/\tau_{fr}|}
P_{||}\  \left\{ 1+    \exp\left[\frac{P_{||}^2}{2mT_{eff}} -
       \frac{\mu_{eff}}{T_{eff}}\right] \right\}^{-1} = 0 ,
\label{e:vls7}
\end{equation}
where for an averaged Yukawa surface term $U$ is only the function of $n$.

Performing products in the first term and taking into account that
$
\vec{P}
\frac{\vec{r} - \vec{r}_0 \tau/\tau_{fr}}{|\vec{r} - \vec{r}_0 \tau/\tau_{fr}|}
$ $= $ $ P_{||}
$
we obtain
$$
g_s\
\left[\kappa
+
\frac{ k\tau_{fr} }{ m \tau }  [ \vec{p} - m \vec{r} / \tau ]
\frac{\vec{r} - \vec{r}_0 \tau/\tau_{fr}}{|\vec{r} - \vec{r}_0 \tau/\tau_{fr}|}
\right]  =  \hspace*{2cm}
$$
$$
g_s\
\left[\kappa
+
\frac{ k\tau_{fr} }{ m \tau }   P_{||}
\frac{|\vec{r} - \vec{r}_0 \tau/\tau_{fr}|}{|\vec{r} - \vec{r}_0
 \tau/\tau_{fr}|}
\right]
=
g_s\
\left[\kappa
+
\frac{ k\tau_{fr} }{ m \tau }   P_{||}
\right] ,
$$
so that Eq. (\ref{e:vls7}) takes the form
\begin{equation}
g_s
\left[\kappa + \frac{ k\tau_{fr} }{ m \tau }  P_{||} \right] -
2\pi  C n_s \frac{\partial U}{\partial n}\ \frac{k\tau_{fr}}{\tau} \
P_{||}
\left\{ 1+    \exp\left[\frac{P_{||}^2}{2mT_{eff}} -
       \frac{\mu_{eff}}{T_{eff}}\right] \right\}^{-1}
\hspace*{-4mm} =0 .
\label{e:vlq6}
\end{equation}

We see that the position of the center of the perturbation, $\vec{r}_0$,
has dropped out.  We indicated this symmetry already in the introduction
when we pointed out that in the spherical scaling expansion all interior
points are equivalent in the sense of their LR features.  We would obviously
get the same result assuming $\vec{r}_0 \equiv 0$ from Eq. (\ref{e:ps0})
on in the course of this derivation.

Dividing both sides by
$(\kappa + \frac{ k\tau_{fr} }{ m \tau }  P_{||} )$
and integrating over $d P_{||}$ leads to
\begin{equation}
1 = 2\pi  C \frac{\partial U}{\partial n}\ \frac{k\tau_{fr}}{\tau}
\hspace*{-2mm}
\int\limits_{-\infty}^\infty
\hspace*{-2mm}
P_{||} d P_{||}
\left[\kappa + \frac{ k\tau_{fr} }{ m \tau }  P_{||} \right]^{-1}
  \left\{ 1+    \exp\left[\frac{P_{||}^2}{2mT_{eff}} -
       \frac{\mu_{fr}}{T_{fr}}\right] \right\}^{-1} \hspace*{-4mm} .
\label{e:vlq7}
\end{equation}
Multiplying both the denominator and the numerator by
$\kappa - \frac{ k\tau_{fr} }{ m \tau }  P_{||} $  and then dropping the
antisymmetric part, which does not contribute to the integral we obtain
\begin{equation}
1 = - 4\pi m  C \frac{\partial U}{\partial n}
\hspace*{-1mm}
\int\limits_0^\infty
\hspace*{-1mm}
d P_{||}   P_{||}^2
\left[\frac{ m^2 \tau^2 \kappa^2}{ k^2\tau_{fr}^2 } -  P_{||}^2 \right]^{-1}
  \left\{ 1+    \exp\left[\frac{P_{||}^2}{2mT_{eff}} -
       \frac{\mu_{fr}}{T_{fr}}\right] \right\}^{-1} \hspace*{-4mm} .
\label{e:vlq8}
\end{equation}
The same way as we got the dispersion relation in the case of plane wave
perturbation from Eq. (\ref{e:vls4}) we get now the relation
\begin{equation}
1 = - 2 \pi m C \sqrt{2mT_{eff}}\ \frac{\partial U}{\partial n}\
\int\limits_0^\infty
  \frac{d y\  \sqrt{y} }{ (y-s) [1+ e^{y-\mu_{fr}/T_{fr}} ] },
\label{e:adis2}
\end{equation}
where $s= m \kappa^2 \tau^2 /( 2 k^2 \tau_{fr}^2 T_{eff}) =
          m \kappa^2 \tau^4 /( 2 k^2 T_{fr} \tau_{fr}^4) $ .

If we have a spherical droplet perturbation of a finite
radius described by Eq.~(\ref{e:ps5})
instead of a cusp, Eq.~(\ref{e:ps0}),
the dispersion relation can be  obtained from
(\ref{e:adis2}) by making the transformation
$
\kappa \longrightarrow \kappa k R^* \tau_{fr} / \tau_{th}
$ arising from comparing the form of the two perturbations (\ref{e:ps0})
and (\ref{e:ps5}).
This leads to
exactly the same equation as the equation
above (\ref{e:adis2}) for the spherical cusp perturbations,
except that in place of $s$ we have $S = m \kappa^2 (R^*)^2 \tau^4
/( 2 T_{fr} \tau_{fr}^2 \tau_{th}^2) $ in the expression.

\section*{\large\bf Appendix C: Spherical droplet with Yukawa forces}
\label{app-yuk}

For the profile (\ref{e:ps5}) the Yukawa force can be written as follows
\begin{equation}
V_{\rm {Yuk}}(r) = {4\pi V_0 \over{ \mu^2 }} \left[
{\mu\over{r}}  e^{-\mu R}
\sinh(\mu r)
\left( {R \over{ \mu }} -
{R \over{ \mu +\alpha }} + {1 \over{ \mu^2 }} -
{1 \over{ (\mu + \alpha)^2 }} \right)
- 1 \right]
\label{e:ay1}
\end{equation}
for $r < R(\tau)$, where $\alpha = k \tau_{fr} / \tau$, and
\begin{eqnarray}
V_{\rm {Yuk}}(r)
&=&
- 4\pi V_0 n_{\rm  c} e^{-\alpha (r-R)} \left[
- {1 \over{ \alpha^2 - \mu^2 }} \right.
- {2\alpha \over{ r (\alpha^2 - \mu^2)^2 }}
\nonumber \\
&+&
{ e^{(\alpha - \mu) (r-R)} \over{ 2 r \mu }}
\left(
  {R \over{ \mu }} - {1 \over{ \mu^2 }}
+ {R \over{ \alpha -\mu }} + {1 \over{ (\alpha - \mu)^2 }}
\right) \\
&+&
\left.
 {e^{(\alpha - \mu) r}\over{2r\mu e^{(\alpha +\mu)R}}} \hspace*{-1pt}
 \left({R \over{\mu}} +
  {1 \over{\mu^2}}
- {R\over{\alpha +\mu}}
- {1 \over{(\alpha + \mu)^2}}
 \right) \hspace*{-2pt}
\right]
\nonumber
\label{e:ay2}
\end{eqnarray}
for $r > R(\tau)$.

The Yukawa energy can be written as
\begin{eqnarray}
E_{\rm {Yuk}}
&=& \int d^3r\ V_{\rm {Yuk}}(\vec{r}) n_1(\vec{r})
\nonumber \\
&=& - {4\pi V_0 \over{ \mu^2 }}
      n_{\rm  c}^2 {4\pi \over 3}
 \left[
    R^3
-   {3 \over 2} R^2
    {\alpha^2 - \alpha \mu -\mu^2 \over{ \mu \alpha (\mu + \alpha) }}
 \right.
\nonumber \\
&+&
  {3 \over 2} R
  \left(
-  {2 \over{ \mu }} {1 \over{ (\mu + \alpha) }}
+  {1 \over{ (\mu + \alpha)^2 }}
+  {1 \over{ \alpha^2 }}
  \right)
\nonumber \\
&+& {3 \over 2}\ \
  \left(
  {(2 \alpha + \mu) \mu \over{ 2 \alpha^3 (\mu + \alpha)^2 }}
- {2 \over{ \mu (\mu + \alpha)^2 }}\
+ {1 \over{ \mu^3 }}
  \right) \\
&-&
  \left.
  {3 \over 2} \mu e^{-2 \mu R}
    \left(
     {R \alpha \over{ \mu (\mu + \alpha) }}
-    {1 \over{ (\mu + \alpha)^2 }}
+    {1 \over{ \mu^2 }}
    \right)^2
  \right] \nonumber
\end{eqnarray}

One sees that substituting expression (\ref{e:ay1}) and (\ref{e:ay2}) into
the Vlasov equation, the perturbation (\ref{e:dis3}) is not a solution of it.
However, for sharply decreasing surfaces $\alpha R > 1$ and
we can consider instead of $V_{\rm Yuk}$ the
average of it.
That is, we use a
surface term, which is given as the average of the Yukawa interaction.
Instead of (\ref{e:ay2}) we introduce for $r > R(\tau)$ (see
Eq.~(\ref{e:3p-1}))
\begin{eqnarray}
\label{e:ayav}
U_{\rm surf} &=& U^0_{\rm surf} n_1(\vec{r})\ =
{4\pi V_0 \over{ \mu^2 }}\  \left[ 1 - \overline{V}_{\rm Yuk} \right] = \\
&=& {4\pi V_0 \over{ \mu^2 }}\  n_1(\vec{r})\
{\alpha \over{ \mu (\mu + \alpha) }}\
 {R + {2\alpha + \mu \over{ 2\alpha (\alpha + \mu) }} \over{
  R^2 + {R \over{ \alpha }} + {1 \over{ 2\alpha^2 }} }}
\nonumber
\end{eqnarray}

The surface energy term in Eq.~(\ref{e:3p-1}) can be written using
(\ref{e:ay2}) and neglecting the terms proportional to $e^{-2\mu R}$
(these terms are small) as
\begin{eqnarray}
\label{e:as1}
\Delta \varepsilon_{\rm surf} \hspace*{-2mm} &=& \hspace*{-2mm}
{4\pi V_0 \over{ \mu^2 }}\ - {E_{\rm Yuk} \over{ \overline{N}_1 }}
\nonumber \\
\hspace*{-2mm} &=& \hspace*{-2mm}
 {1 \over{ \overline{N}_1 }} {4\pi V_0 \over{ \mu^2 }}\ \left[
  {R^2 \alpha \over{ 2\mu (\mu + \alpha) }} + {R \over{ (\mu + \alpha)^2 }}
  {2\alpha + \mu \over{ 2\mu }} + {4\alpha + \mu \over{
   4\alpha (\mu + \alpha)^2 }} -{1 \over{ 2 \mu^3 }}  \right]
\nonumber \\
\hspace*{-2mm} &\to& \hspace*{-2mm} {6\pi V_0 \over{ \mu^3 R }}\
{\alpha \over{ \mu + \alpha }}
\quad \mbox{for }\ \alpha R \gg 1\ .
\end{eqnarray}

\section*{\large\bf Appendix D: The mechanical instability region}
\label {sec-isoth}

The total energy density of the system can be written as
$e_{{pot}}(n)
+e_{F}$, with the potential energy density $e_{{pot}}(n)$ and the
kinetic (Fermi)
energy density $e_{F}$. The latter can be written as $e_{F}=const.\ T^{5/2}
F_{3/2}(\mu/T)$, using the integrals $F_{i/2}(\eta) = \int\limits_0^{\infty}\
dx
{x^{i/2} \over{ 1 + \exp(x-\eta) }}$. The density can be expressed as
$n=const.\ T^{3/2} F_{1/2}(\mu/T)$. Isotherm expansion, $dT=0$, leads  to
the change of Fermi energy
$$
de_{F} = 3 T {F_{1/2}(\mu/T) \over{ F_{-1/2}(\mu/T) }} dn
$$
The pressure should be calculated from the free energy density
$f(T,n)=e-Ts$, with
the entropy density $s={5\over 3} e_{F}/T - n \mu/T$:
$$
p = n^2 {\partial f/n \over{ \partial n }} = n
{\partial e_{{pot}}(n) \over{ \partial n }} -
e_{{pot}}(n) + {2\over 3} e_{F}
$$
The region of mechanical instability where the derivative of the pressure above
with respect to the density at constant temperature is negative:
$$
{d p\over{ d n }} = n {\partial^2 e_{{pot}}(n) \over{
\partial n^2 }} + 2 T {F_{1/2}(\mu/T) \over{ F_{-1/2}(\mu/T) }}
= n {\partial U \over{ \partial n }} + 2 T {F_{1/2}(\mu/T) \over{
F_{-1/2}(\mu/T) }}
$$
The onset of the instability is determined then by
\begin{equation}
1 = - {1\over 2} c T^{1/2}
{\partial U \over{ \partial n }} F_{-1/2}(\mu/T) \quad , \quad
c = {g \over{ 4 \pi^2 }} \left( {2m \over { \hbar^2 }} \right)^{3/2}
\end{equation}

The condition of the dynamical instability derived in Appendix B is
$$
1 = - {2\pi \tau_{fr} m \over{ \tau }} {g \over{ (2\pi \hbar )^3 }}
\sqrt{2 m T_{fr}}
{\partial U \over{ \partial n }} \int\limits_0^{\infty} {dy \sqrt{y}
\over{
(y-s) \left( 1 + \exp(y-\mu_{{eff}}/T_{{eff}}) \right) }}
$$
substituting $\tau_{fr}/\tau=\left( T/T_{fr} \right)^{1/2}$ for the
critical mode ($s=0$ growing) one gets
\begin{equation}
1 = - {1\over 2} \left( {2m \over { \hbar^2 }} \right)^{3/2} {g \over{
4 \pi^2 }} T^{1/2}
{\partial U \over{ \partial n }}
F_{-1/2}(\mu_{{eff}}/T_{{eff}}) \quad ,
\end{equation}
which is the same as the condition for the isothermal spinodal.

\section*{\large\bf Appendix E: Critical radius}

Let us introduce the notation
$
\langle n_1^2 \rangle\ \frac{4\pi}{3}\ R^3 \equiv \overline{N}_1
$
and rewrite Eq.~(\ref{e:as1}) as
\begin{equation}
\Delta E_{\rm {surf}} =
{8\pi^2 V_0 \over{ \mu^2 }}  {\alpha \langle n_1^2
\rangle \over{ R \mu (\mu +\alpha) }}
R^2 \equiv a(\tau) R^2 .
\label{e33d}
\end{equation}
Using Eq.~(\ref{e:3p-4}) the total energy change due to the formation of a
droplet of size $R$ can be cast in the form
$$
\Delta E (R) =  \Delta E_{\rm {kin}}
- \underbrace{\left(
   {\beta \over{3}}
 + {4\pi V_0 \over{3 \mu^2 }}
 - { (\sigma+1) (\sigma+2) \over 2 } \gamma n_0^{\sigma}
  \right)
4 \pi \langle n_1^2 \rangle }_{\equiv b} R^3
+ a R^2.
$$
Thus, assuming that the contribution of kinetic energy is independent of $R$
we obtain the critical radius from $\partial \Delta E / \partial R = 0$
$$
R^*_{\rm crit} = \frac{2 a(\tau)}{ 3 b} .
$$
At $R^*_{\rm crit}$ the function $\Delta E(R) $ has its maximum, thus
bubbles or droplets smaller than $R^*_{\rm crit}$ shrink while larger
than $R^*_{\rm crit}$
grow.

\section*{\large\bf Acknowledgments}

We thank J. Bondorf, I. Mishustin,
W. N\"orenberg
for stimulating discussions.
This work was supported in
part by the Norwegian Research Council (NFR), the Hungarian
Research Foundation OTKA and part by Contract N$^o$ ERB-CIPA-CT92-4023.
One of the authors (J.N) would like to express her thanks to Prof. W.
Greiner and
the University of Frankfurt for their kind hospitality, where part of this work
was done.

\section*{\large\bf Notes}

E-mail: L. P. Csernai: csernai@fi.uib.no, http://www.fi.uib.no/\verb+~+csernai;
Judit N\'e\-meth: judit@hal9000.elte.hu;
G. Papp: G.Papp@gsi.de http://www.gsi.de/\verb+~+papp.\\
$^\dagger$: Permanent address

\newpage


\begin{thebibliography}{99}
\bibitem{csm95}
L.P. Csernai and I.N. Mishustin, {\it Phys. Rev. Lett.}
          {\bf 74} (1995) 5005.

\bibitem{csmm95}
L.P. Csernai, I.N. Mishustin and \'A. M\'ocsy,
          {\it Heavy Ion Phys.} {\bf    } (1995) in preparation.

\bibitem{kn:be88}
G.F. Bertsch and S. Das Gupta, {\it Phys. Rep.}
          {\bf 160} (1988) 189.

\bibitem{vv94} R. Venugopalan and A.P. Vischer, {\it Phys. Rev.} {\bf E49}
               (1994) 5849.

\bibitem{ck92}
L.P. Csernai and J.I. Kapusta, {\it Phys. Rev. Lett.}
            {\bf 69} (1992) 737; {\it Phys. Rev.} {\bf D46} (1992) 1379.

\bibitem{ckz93}
L. P. Csernai, et al.,  Z. Phys. {\bf C58}, 453 (1993).

\bibitem{tl80} L.A. Turski and J.S. Langer, {\it Phys. Rev.} {\bf A22}
                (1980) 2189.

\bibitem{lt73} J.S. Langer and L.A. Turski, {\it Phys. Rev.} {\bf A8}
                (1973) 3230.

\bibitem{np90}
J. N\'emeth, G. Papp, C. Ngo and M. Barranco,
       {\it Phys. Scripta} {\bf T32} (1990) 160.

\bibitem{cc94}
T. Cs\"org\H o and  L.P. Csernai, {\it Phys. Lett.}
               {\bf B333} (1994) 494.

\bibitem{aram}
J. Kapusta and  A. Mekjian, {\it Phys. Rev.} {\bf D33} (1986) 1304.

\bibitem{kn:rand}
M. Colonna, Ph. Chomaz, J. Randrup, {\it Nucl. Phys.} {\bf A567} (1994)
637.

\bibitem{kn:mis} A.B. Larionov and  I. Mistushin, {\it Sov. J. Nucl. Phys.}
            {\bf 57}  (1994) 636.

\bibitem{pnb92}
G. Papp, J. N\'emeth, and J. Bondorf, {\it Phys. Lett.} {\bf B278} (1992) 7.

\bibitem{pw95}
G. Papp, W. N\"orenberg, GSI Preprint 95-30, submitted to
{\it Heavy Ion Phys.} (1995)
\end{thebibliography}
\end{document}